\documentclass[num-refs]{wiley-article}

\papertype{}

\pdfoutput=1

\usepackage[utf8]{inputenc}
\usepackage{caption}
\usepackage{subcaption}
\usepackage{xspace}
\usepackage{tabularx}
\usepackage{enumitem}
\usepackage[bookmarks=false]{hyperref}
\usepackage{xcolor}
\usepackage{balance}
\usepackage{tikz}
\usepackage{framed}
\usepackage{pgfplots}
\usepackage{listings}
\usepackage{multirow}

\definecolor{transfertoserver}{HTML}{D7191C}
\definecolor{database}{HTML}{FDAE61}
\definecolor{transfertoclient}{HTML}{ABDDA4}
\definecolor{rendering}{HTML}{2B83BA}

\usepackage{algorithm}
\usepackage{algorithmic}

\pgfplotsset{compat=1.14} 

\newcommand{\question}[2]{{\bf RQ#1. #2}}
\newcommand{\answer}[2]{\vspace{.3cm}{\centering\fbox{\parbox{0.98\columnwidth}{\textbf{Answer to RQ#1}. #2}}}\vspace{.3cm}}

\newcommand{\mycode}[1]{{\small \texttt{#1}}\xspace}
\newcommand{\tabincell}[2]{\begin{tabular}{@{}#1@{}}#2\end{tabular}}

\newcommand\review[1]{{{blue}{#1}}}

\AtBeginDocument{%
}
\makeatletter
\newcommand{\ALC@uniqueautorefname}{Line}
\makeatother

\newcommand{\trace}{web trace\xspace}
\newcommand{\tool}[0]{BikiniProxy\xspace}
\newcommand{\extension}[0]{BugBlock\xspace}
\newcommand{\bench}[0]{DeadClick\xspace}

\newcommand{\nbBug}[0]{4282\xspace}
\newcommand{\nbReproducedBug}[0]{555\xspace}
\newcommand{\visitedPage}[0]{96174\xspace}

\begin{document}

\title{Fully Automated HTML and Javascript Rewriting for Constructing a Self-healing Web Proxy}

\author[1]{Thomas Durieux}
\author[2]{Youssef Hamadi}
\author[3]{Martin Monperrus}

\affil[1]{INESC-ID and IST, University of Lisbon, Portugal}
\affil[2]{Uber Elevate Research, Paris, France}
\affil[3]{KTH Royal Institute of Technology Stockholm, Sweden}

\corraddress{Thomas Durieux, INESC-ID, and IST, University of Lisbon, Portugal}
\corremail{thomas@durieux.me}

\fundinginfo{This material is based upon work supported by Funda\c{c}\~ao para a Ci\^encia e a Tecnologia (FCT), with the reference PTDC/CCI-COM/29300/2017.}


\maketitle

\begin{abstract}
Over the last few years, the complexity of web applications has increased to provide more dynamic web applications to users.
The drawback of this complexity is the growing number of errors in the front-end applications. 
In this paper, we present an approach to provide self-healing for the web.
We implemented this approach in two different tools:
1) \tool, an HTTP repair proxy, and 2) \extension, a browser extension.
They use five self-healing strategies to rewrite the buggy HTML and Javascript code to handle errors in web pages.
We evaluate \tool and \extension with a new benchmark of \nbReproducedBug reproducible Javascript errors of which 31.76\% can be automatically self-healed by \tool and 15.67\% by \extension.
\end{abstract}

\section{Introduction}\label{sec:introduction}

According to \cite{rankinglib}, at least 76\% of all websites on the Internet use Javascript.
The Javascript code used in today's web pages is essential: it is used for social media interaction, dynamic user-interface, usage monitoring, advertisement, content recommendation, fingerprinting, etc, all of this being entirely part of the ``web experience''.
For example, when a user browses the website \url{cnn.com}, she is loading more than 125 Javascript files, which represent a total of 2.8 megabytes of code.

The drawback of this complexity is the growing number of errors in web pages.
For instance, a common Javascript error is due to uninitialized errors, resulting in an error message such as \texttt{cannot read property X of null}. Ocariza et al. \cite{ocariza2011Javascript} have performed a systematic study showing that the majority of the most visited websites contain Javascript errors.

In this paper, we propose a novel technique to provide self-healing for the web. 
It is along the line of previous work on self-healing software  \cite{keromytis2007characterizing,koopman2003elements}, also called failure-oblivious computing (e.g. \cite{rinard2004enhancing,durieux2017dynamic}), automated recovery and remediation (e.g. \cite{candea2003jagr}).
The majority of the self-healing literature focuses on the C/Unix runtime.
On the contrary, we are interested in the Javascript/browser runtime, which is arguably much different.
Indeed, the topic of self-healing for the web is a very little researched area \cite{carzaniga2010automatic,carzaniga2015automatic}. 

Our novel self-healing technique is founded on two insights.
Our first key insight is that {proxying the source code before it is executed can be used for providing self-healing capabilities to an application. In this paper, we focus on two types of proxy: 1) an HTTP proxy between the browser and the web server, 2) a browser extension that modifies the request inside the browser.}
Our second key insight is that the most common Javascript errors can be fixed by an automated rewriting of HTML or Javascript code.

In this paper, we present two novel self-healing tools: a proxy for the web called \tool, and a browser extension called \extension.
The two tools contain five self-healing strategies that are specifically designed for Javascript errors.
Those strategies are based on rewriting, defined as an automated modification of the code. 
\tool and \extension automatically modify the Document Object Model (DOM) of HTML pages or automatically transforms Javascript abstract syntax trees (AST).

Our approach does not make any assumption on the architecture or libraries of web applications.
First, proxy servers are used in most web architectures, and browser extensions are commonly used to change the browser behavior, for example, to block ads.
Second, our approach does not require a single change to existing web pages and applications.
As such, it is highly applicable.

We evaluate our approach as follows.
First, we set up a crawler to randomly browse the Internet, for each browsed page, it logs the Javascript errors, if any, occurring during the loading of the page content.
Second, we observe how \tool and \extension heal those errors by collecting and comparing traces.
Over eight full days, our crawler has visited 96174 web pages and identified 4282 web pages with errors. We observed that 3727 errors were either transient (due to asynchronicity \cite{ocariza2011Javascript}) or fixed by the developer after crawling. 
Eventually, we evaluated \tool and \extension on \nbReproducedBug web pages with errors, representing a random sample of real field errors.
\tool is able to make all errors disappear in 176/\nbReproducedBug of the cases, that is 31.76\%.
\extension is able to make all errors disappear in 87/\nbReproducedBug of the cases, that is 15.67\%.
In the best cases, the self-healing provides the user with new features or new content. We provide a detailed qualitative and quantitative analysis of the main categories of self-healing outcome.
{
To sum up, \tool and \extension are novel fully automated self-healing techniques that are designed for the web, evaluated on \nbReproducedBug real Javascript errors, and based on original self-healing rewriting strategies for HTML and Javascript.}

Our contributions are:
\vspace{-1.5em}
\begin{itemize}
\item A novel self-healing approach for today's web: implemented into two independent tools: a self-healing proxy (\tool) and a self-healing browser extension (\extension).

\item Five self-healing strategies for the web, specifically designed to automatically recover from the most frequent Javascript errors happening in the browser.

\item A benchmark of \nbReproducedBug real web pages with Javascript errors. Special care has been taken so that all error are fully reproducible for future experiments in this research area.

\item An evaluation of \tool and \extension over \nbReproducedBug Javascript errors from our benchmark.
It shows that \tool makes all errors disappear for 31.76\% of web pages with errors {and \extension handles all the errors in 15.67\% of the web pages from our benchmark}. 
This quantitative evaluation is complemented by a  qualitative analysis of \tool's and \extension's effectiveness.

\item The implementations of \tool and \extension are open-source and publicly available for future research in \cite{repo}.

\end{itemize}

{
This paper is an extended version of \cite{durieux2018fully}. In the conference paper, we presented only \tool and its evaluation. In this extension, we devise and evaluate the corresponding browser extension \extension. Both tools have the same purpose and target the same failures, but they target different use cases and execution environments. 
Having both shows that our underlying self-healing concept is generic and widely applicable.}

The remainder of this paper is organized as follows.
\autoref{sec:background} explains the background of \tool.
\autoref{sec:contribution} details our approach for introducing self-healing web application and two implementations: \tool and \extension.
\autoref{sec:evaluation} details the evaluation.
{\autoref{sec:discussion} discusses the security and applicability aspects of the approaches.}
\autoref{sec:threats_validity} details the threats to validity of the contribution.
\autoref{sec:rw} presents the related works, and \autoref{sec:conclusion} concludes.

\section{Background}\label{sec:background}

\subsection{The Complexity of Today's Web}

A web page today is a complex computational object.
A modern web page executes code and depends on many different resources.
Those resources range from CSS styles of thousands of lines, external fonts, media objects, and last but not least, Javascript code.

For example, when a user browses the website \url{cnn.com}, he is loading more than 400 resources, and 125 of them are Javascript, which represents a total of 2.8mb of code.
Anecdotally, back in 2010, the same web page \url{cnn.com} contained 890kb of Javascript code \cite{ocariza2011Javascript} (68\% less code!).

Today's web page Javascript is essential: it is used for social media interaction, usage monitoring, advertisement, content recommendation, fingerprinting, etc., all of this being entirely part of the ``web experience''.
Consequently, 76\% of all websites on the Internet use Javascript \cite{rankinglib}.
To this extent, a web page today is a program, and as such, suffers from errors.

\subsection{Javascript Errors}
Web pages and applications load and execute a lot of Javascript code \cite{ocariza2017study}.
This code can be buggy; in fact, the top 100 of the most visited websites contains Javascript errors \cite{ocariza2011Javascript}.
One kind of Javascript error is an uncaught exception, which is similar to uncaught exceptions in modern runtimes (Java, C\#, Python).
{While the Javscript community uses the term ``error'', the research community use ``failure'' in this case.}
Those errors are thrown during execution if the browser state is invalid as when accessing a property on a null element (a null dereference). 

If an error is not caught by the developer, the execution of the current script is stopped. 
In Javascript, there is a different ``execution scope'' for each loaded scripts (i.e., for each HTML script tag) and for each asynchronous call.
Consequently, contrary to classical sequential execution, in a browser, only the current execution scope is stopped, and the main thread continues running.
This means that one can observe several uncaught exceptions for a single page.
{Therefore, depending on where the error is happening the error is perceived differently by the end-user. 
If the error happens in the main execution scope, the failure will prevent the execution of all the code that is after the crash. This will have an important impact on the user experience. 
However, if the error happens is an execution scope that is not meant to provide a feature for the user, the user will not perceive it, and it will give the impression that the browser tolerates the failure. But, in this case, there can still be an observable impact on the system, for example, if the error happens in the logging module.}

The uncaught errors are logged in the browser console that is accessible with the developer tool.
Most browsers provide an API to access all the errors that are logged in the browser console. 
This means that it is relatively easy to monitor Javascript errors in web applications.

\subsection{Web Proxies}
A web proxy is an intermediate component between a client browser and web server. 
In essence, a proxy transmits the client request to the server, receives the response from the server, and forwards it to the client browser.
On the web, proxies are massively used for different purposes.

\begin{enumerate}
\item A \texttt{Network Proxy} is used to expose a service that is not directly accessible because of network restrictions \cite{luotonen1998web}.
The proxy, in this case, is a bridge between two networks, and its only task is to redirect requests and responses.
For example, a popular network proxy is Nginx. It is used to expose websites on port 80 which are indeed served on other ports $>1024$. This avoids granting root access to web applications.
\item A \texttt{Cache Proxy} is a proxy that is used to cache some resources directly in the proxy in order to improve the serving time to an external resource \cite{pistriotto2000method}. 
The cache proxy stores the response of the server locally, and if a request is made for the same resource, the local version is directly sent to the client without sending the request to the server.
A \review{widely} used cache proxy is, for example, a content-delivery network (CDN) that provides optimized access to static resources on the Internet.
\item A \texttt{Security Proxy} is used to verify whether a client browser is legitimate to access a server \cite{krueger2010tokdoc}.
This type of proxy can be used, for example, to protect a server against Denial-of-service attacks.
\item A \texttt{Load-balancer Proxy} is used on popular applications to distribute the load of users on different backend servers \cite{bowman2003load}.
A load balancer can be as simple as a round-robin, but can also be more sophisticated. For instance, a load-balancer can try to find the least loaded server available in the pool.
\end{enumerate}

\subsection{Browser Extensions}

Browser extensions are used by end-users to extend the functionalities of their browsers.
It is used for changing the user interface of the browser, for blocking ads, and for increasing the privacy.

The browser extensions are HTML/Javascript micro-applications that have more permissions than the traditional browser Javascript.
Those applications work at three different levels in the browser.
The first level is called ``background''. It means that the code in the extension constantly runs, and is generally used for the core behavior of the extension and the state storage. It is also the only part of the extension that can register to specific events, such as a new request event, a new tab event.
The second level is the ``action'' level. It is used for providing information to the user. It is generally presented as an icon on the right of the URL bar. This level has access to the state of the first level.
The third level is the ``injected script'' level. This level is used to inject specific Javascript behavior on the page.
It can be used, for example, by a password manager to inject the login and password in a login form.

With those three levels, browser extensions have the possibility to drastically change the behavior of web pages and interact with the users.
One way to modify the behavior of web pages is to block, redirect, and inject scripts in web pages.
For example, ads blockers block the HTTP requests that match specific patterns.

\section{Fully Automated Self-healing Approach and Tools}\label{sec:contribution}

{We now present our novel approach to fully automate self-healing of HTML and Javascript in production.
This approach is composed of three main parts:
firstly, the failure-oblivious computing central component (see \autoref{sec:failure-oblivious}),
secondly, the self-healing strategies that are used to inject failure-oblivious properties to an application (see \autoref{sec:repair_models}),
and finally, the two implementations of the approach \tool, a HTTP-proxy (see \autoref{sec:architecture}), \extension, a browser extension (see \autoref{sec:extension}).

Those two different implementations show the power of our self-healing concept in different environments.
The first one is more powerful but less practical (\tool), and the second one is less powerful but more practical (\extension).}

\subsection{Failure-oblivious Computing}\label{sec:failure-oblivious}

Our work is founded on the failure-oblivious computing principle \cite{rinard2004enhancing}.

\begin{definition}[Failure-oblivious computing principle]
\label{def:failure-principle}
It is possible to perform speculative execution after the failure point instead of crashing. 
\end{definition}

{The goal of failure-oblivious computing is to provide a degraded behavior instead of no behavior at all when the application is crashing. 
It is not a long-term solution, but it increases availability.
It can be compared to a temporary hotfix to mitigate a problem before it is permanently fixed.}
Failure-oblivious computing is desirable when availability must be maximized while having limited resources for engineering expensive highly reliable and available software. There is indeed an engineering tradeoff between cost and reliability \cite{CostEffectiveSelfHealing2010}.
Failure-oblivious computing fits in the general context of self-healing software \cite{keromytis2007characterizing,koopman2003elements}, that is software with seatbelts and airbags \cite{Berger2012}. Self-healing is also known as runtime repair \cite{lewis2010runtime} or automated recovery \cite{dagstuhlSelfHealing,gu2016automatic}.

\subsection{Self-healing Strategies}\label{sec:repair_models}

{This section presents the five self-healing strategies that we designed and implemented to mitigate Javascript errors.
We designed those strategy to target the most frequent Javascript errors happening on the web (see \autoref{sec:benchmark} for how we identify those most common errors).
Those fives strategies are implemented in two tools: in \tool and in \extension.}

We designed: \vspace{-1.5em}
\begin{enumerate}
\item HTTP/HTTPS Redirector that changes HTTP URLs to HTTPS URLs (see \autoref{sec:strateogy:https}).
\item HTML Element Creator that creates missing HTML elements (see \autoref{sec:strateogy:html}).
\item Library Injector injects missing libraries in the page (see \autoref{sec:strateogy:inject}).
\item Line Skipper wraps a statement with an if to prevent invalid object access (see \autoref{sec:strateogy:skip}).
\item Object Creator initializes a variable with an empty object to prevent further null dereferences (see \autoref{sec:strateogy:create}).
\end{enumerate}

\subsubsection{HTTP/HTTPS Redirector}\label{sec:strateogy:https}
The modern browsers have the policy to block unsecured content (i.e., HTTP resources) in secured web pages (HTTPS).
For example, \texttt{https://foo.com} cannot load \texttt{http://foo.com/f.js} (no https).
The rationale is that the unsecured requests can be easily intercepted and modified to inject scripts that will steal information from the secure page, for example, your banking information.

As of 2019, we still are in a period of transition where both HTTP and HTTPS web pages exist, and some websites provide access to their content with both HTTP and HTTPS protocol.
Consequently, it happens that the developers forget to change some URL in their HTTPS version, and those resources are blocked, resulting in incomplete web pages or Javascript errors.

The self-healing strategy of HTTP/HTTPS Redirector is to change all the HTTP URL by HTTPS URL in HTTPS pages.
By doing this, all resources are loaded, and the unwanted behavior due to blocked resources is fixed.

\subsubsection{HTML Element Creator}\label{sec:strateogy:html}

As shown by Ocariza et al. \cite{ocariza2017study}, most of the Javascript errors are related to the DOM.
This is especially true when the developers try to dynamically change the content of a specific HTML element using getElmentById, like:
\begin{lstlisting}
document.getElmentById(``elementID'').innerText = ``Dynamic content'';
\end{lstlisting}
Since the HTML and the Javascript are provided in different files, it is not rare that an HTML element with an ID is removed or changed without changing the associated Javascript code.
For example, if DOM element \mycode{"elementID"} is removed, \mycode{document.getElmentById(...)} returns \mycode{null} and the execution results on a null dereference when the property \mycode{innerText} is set.

When an error happens, \tool and \extension determine if the error is related to the access of a missing element in the DOM: it does so by looking at the Javascript code at the failure point.
If it is the case, ``HTML Element Creator'' extracts the query that the Javascript used to access the element and create an empty and invisible HTML element in the DOM.
The Javascript code then runs without error, and the execution continues without affecting the client browser.

\subsubsection{Library Injector}\label{sec:strateogy:inject}
In Javascript, it is a common practice to rely on external libraries to facilitate the development of the web applications.
Some of these libraries are extremely popular and are used by millions of users every day, like jQuery or AngularJS.
Sometimes, these libraries are not correctly loaded into web pages, and this produces a very characteristic error:
for example, \mycode{jQuery is not defined}. In those cases, we can parse the error message and determine which library is missing.

To do so, we realize an initial offline training phase, missing libraries are simulated on a test website,  and reference errors are collected for the top 10 libraries presented in \cite{rankinglib}.
Based on these reference errors, error parsing rules have been manually written to determine which library is missing.
When Library Injector detects that a web page contains an error related to a missing library, it injects the related library on the web page.
The rewritten page contains the missing library, and the web page can be completely loaded.

\begin{algorithm}[t]
  \begin{algorithmic}[1]
    \REQUIRE{E: error}
    \REQUIRE{R: resource}
    \STATE{(line, column, resource\_url) = extractFailurePoint(E, R.body)}
    \IF{resource\_url is R.url}
        \STATE{ast = getAST(R.body)}
        \STATE{elem = getElementFromAst(ast, line, column)}
        \STATE{wrapElemWithIf(elem)}
        \STATE{R.body = writeAST(ast)}
    \ENDIF
  \end{algorithmic}
  \caption{Algorithm to rewrite Javascript code with ``Line Skipper'' strategy}
  \label{algo:line_skipper}
\end{algorithm}

\subsubsection{Line Skipper}\label{sec:strateogy:skip}

The errors \mycode{XXX is not defined} and \mycode{XXX is not a function} in Javascript are errors that are related to invalid access to a variable or a property.
\mycode{XXX is not defined} is triggered when an identifier (name of variable/function) is used but was never defined in the current scope.
For example, if we call \mycode{if(m)\{\}} without defining \mycode{m}, the execution ends with the error \mycode{`m' is not defined}.

The second error, \mycode{XXX is not a function}, is triggered when a variable that is not a function is called.
For example, the code \mycode{var func = null; func()} will trigger the error \mycode{func is not a function}.

To avoid these errors, Line Skipper wraps the statement that contains the invalid code with an if that verifies that the element is correctly defined for the first error, \mycode{if (typeof m != `undefined' \&\& m) \{if(m)\{\}\}}.
And to verify that a variable contains a reference to  function, the rewriter changes the Javascript code as follows \mycode{if (typeof func === `function') \{func()\}}.

\autoref{algo:line_skipper} presents the algorithm of Line Skipper.
First Line Skipper extracts the line, column, and URL of the resource from the failure point of the error.
Line Skipper verifies that the current request is the resource that contains the error.
Then Line Skipper extracts the AST from the Javascript code and looks for the element that is not defined.
Finally, the AST is transformed back to a textual form to be sent back to the client.

\subsubsection{Object Creator}\label{sec:strateogy:create}

{The Object Creator strategy is created to handle null dereferences (e.g., \texttt{NullPointerException} in Java), which is one of the most frequent error type \cite{toperror} in Java and also the most frequent failure in Javascript \cite{ocariza2017study}.
The typical strategy to handle those errors is to initialize the null variable or replace the null expression by another non-null variable or expression.
Since Javascript is an untyped language, all null dereferences can be handled by initializing the null variable/expression with a generic empty object \mycode{var obj = {};}.
For example, the code \mycode{var m = null; m.test = '';} will trigger the error \mycode{Cannot set property test of null} and can be handled by adding \mycode{if (m == null) \{m = \{\};\}} before setting \mycode{m.test}.}

\subsection{\tool}\label{sec:architecture}

\begin{figure}[t!]
\centering
\includegraphics[width=0.95\columnwidth]{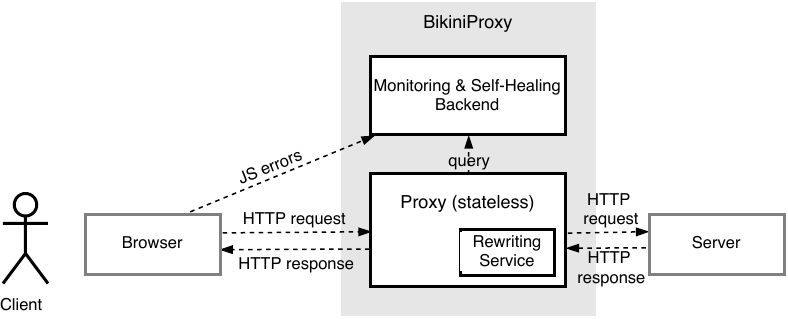}
\caption{The Architecture of \tool. The key idea is that all requests are proxied by ``\tool''. Then, if an error is detected, a self-healing strategy based on HTML and/or Javascript rewriting is automatically applied.}

\label{fig:architecture}
\end{figure}

We now are presenting \tool, the first of the two implementations of our approach.
The intuition behind \tool is that a proxy between web applications and the end-users could provide the required monitoring and intercession interface for automatic error handling.
This is the concept of ``self-healing proxy'' that we explore in this paper.  {\tool is composed of three main parts that are presented in 
\autoref{fig:architecture}.}
\begin{enumerate}
\item Proxy (see \autoref{sec:proxy}) is a stateless HTTP proxy that intercepts the HTML and Javascript requests between the browser (also called ``client'' in this paper) and the webserver.
\item The Rewriting Service (see \autoref{sec:rewriting_service}) that contains the self-healing strategies to handle Javascript errors. 
\item The Monitoring and Self-healing Backend (see \autoref{sec:backend}) stores information about the known errors that have happened and the success statistic of each self-healing strategies for each error.
\end{enumerate}

Let us start with a concrete example.
Bob, a user of the website \url{http://foo.com} browses the page \mycode{gallery.html} and uses \tool to improve his web experience.
Since \tool is a proxy, when Bob opens \mycode{gallery.html}, the request goes through the proxy.
When the request is made, \tool queries the backend to know whether another, say Alice, has experienced errors on \mycode{gallery.html}.
Indeed, Alice's browser got a \mycode{jQuery is not defined} error two days before.
The backend sends this error to the proxy, which consequently launches the Rewriting Service to handle the error.
For \mycode{gallery.html}, the rewriting is HTML-based and consists of injecting the library jQuery in the HTML response.
\tool also injects its error monitoring framework before sending the rewritten response to Bob's browser.
The rewritten page is executed by Bob's browser, \tool's monitoring tells the proxy that Alice's error does not appear anymore, meaning that {
the self-healing strategy handled it}.

\autoref{algo:main} shows the complete workflow of \tool.
\tool receives the HTTP request from the browser (\autoref{algo:bikiniproxy:request}).
Then it redirects the request to the Web Server (\autoref{algo:bikiniproxy:response}) like any proxy.
For each HTML response, \tool injects a framework (\autoref{algo:bikiniproxy:inject}) to monitor the Javascript errors in the client browser.
When an error happens on the client browser, it is sent to \tool's backend for being saved in a database.

{\tool queries the Monitoring \& Self-healing backend to know which Javascript resource has thrown an error in the past}:
for each HTML and Javascript resource, \tool queries the backend service with the URL of the requested resource to list all the known errors (\autoref{algo:bikiniproxy:errors}).
If there is at least one known error, \tool triggers the Rewriting Service to apply the self-healing strategies the requested resource (\autoref{algo:bikiniproxy:rewrite}).
Then the response is sent to the client (\autoref{algo:bikiniproxy:send}) with a unique id to monitor the effectiveness of the applied self-healing strategy.

\begin{definition}[Resource]
\label{def:resource}
A web \textbf{resource} is a content on which a web page is dependent. For instance, an image or a Javascript script is a web resource. 
In this paper, a web resource is defined by
1) an URL to address the resource;
2) its content (text or binary content)
and 
3) the HTTP headers that are used to serve the resource.
The resource can be used as an attribute of an HTML tag (\textless script\textgreater, \textless img\textgreater, \textless link \textgreater, \textless iframe\textgreater, etc.) or used as an AJAX content.\footnote{AJAX means requested programmatically in Javascript code}
\end{definition}

\begin{algorithm}[t]
  \begin{algorithmic}[1]
    \REQUIRE{B: the client browser}
    \REQUIRE{W: the Web Server}
    \REQUIRE{R: the rewriting services}
    \REQUIRE{D: \tool Backend}
    \WHILE{new HTTP $request$ from B}\label{algo:bikiniproxy:request}
        \STATE{$response$ $\leftarrow$ W($request$)} \label{algo:bikiniproxy:response}
        \IF{request is html page}
            \STATE{$response\leftarrow inject\_bikiniproxy\_code(response)$}\label{algo:bikiniproxy:inject}
        \ENDIF
        \STATE{$errors$ $\leftarrow$ $D.previous\_errors\_from(request_{url})$}\label{algo:bikiniproxy:errors}
        \IF{errors is not empty}
            \FOR{$r$ in $R$}
                \IF{$isToApply(r, errors, request, response)$}
                    \STATE{$response \leftarrow r.rewrite(response, request, errors)$}\label{algo:bikiniproxy:rewrite}
                    \STATE{$response \leftarrow response + uuid$}   
                \ENDIF
            \ENDFOR
        \ENDIF
        \STATE{$send(response)$}\label{algo:bikiniproxy:send}
    \ENDWHILE
  \end{algorithmic}
  \caption{The main \tool algorithm}
  \label{algo:main}
\end{algorithm}

\subsubsection{The Proxy}\label{sec:proxy}

A proxy intercepts the HTML code and the Javascript code that is sent by the webserver to the client browser.
By intercepting this content, the proxy can modify the source code of the website and therefore change the behavior of the web application.
One well-known example of such a change is to minimize the HTML and Javascript code to increase the download speed.

In \tool, the proxy automatically changes the Javascript code of the web application to handle known errors.
\tool is configured with what we call ``self-healing strategies''. A self-healing strategy is a way to recover from a certain class of errors automatically. The strategies are presented in \autoref{sec:repair_models} and how they are applied is presented in \autoref{sec:rewriting_service}.

\subsubsection{Rewriting Service}\label{sec:rewriting_service}

The role of the Rewriting Service is to rewrite the content of the Javascript and HTML resources in order to:
1) monitor the Javascript errors that happen in the field
2) change the behavior when a Javascript resource has been involved in an error in the past.
In this paper, a ``known error'' is an error that has been thrown in the browser of a previous client, that has been detected by the monitoring feature of \tool and that has been saved in the Monitoring \& Self-healing Backend (see \autoref{sec:backend}).

We design five self-healing strategies that target the most frequent Javascript errors that we observe when we craw the Internet. Those strategies are presented in \autoref{sec:repair_models}.
In addition, the Rewriting Service is plugin-based, it can be easily extended with new self-healing strategies to follow the fast evolution of the web environment.

\subsubsection{Monitoring and Self-healing Backend}\label{sec:backend}

The Monitoring and Self-healing Backend fulfills three tasks.
The first task is to receive and store all the Javascript errors happening on client browsers.
The Backend provides an API for \tool to query if a specific resource (URL) contains known errors.

The second task of the backend is to monitor the effectiveness of the different self-healing strategies. 
Each time that the section of the Javascript code rewrite by one of the five self-healing strategies is executed, an event is sent to the \tool backend to keep track of the activation of the different strategies.
Based on the number of activation and the number of errors per page, we can estimate the relative effectiveness of the self-healing strategies.
    
The third task is to provide a layer of communication with the developers about the monitored errors and the effectiveness of all self-healing strategies.
For example, the following message can be given to the developer: ``The strategy \texttt{Library Injector} has injected jQuery 22 times in the page \mycode{gallery.html} to handle the error \mycode{jQuery is not defined}''.
This is valuable information to assist the developers in designing a permanent fix.
The backend also provides a visual interface that lists all the errors that the end-users face during the browsing of the web page.

\subsection{\extension}\label{sec:architecture_extension}

{We now present the components of the second implementation of our approach.}
\extension is the second implementation of our approach that aims to provide self-healing abilities to web applications.
{
\extension is like an ad-blocker, but instead of blocking the advertisement, it blocks the Javascript errors.}
This implementation is independent of \tool.

\begin{figure}[t!]
\centering
\includegraphics[width=0.75\columnwidth]{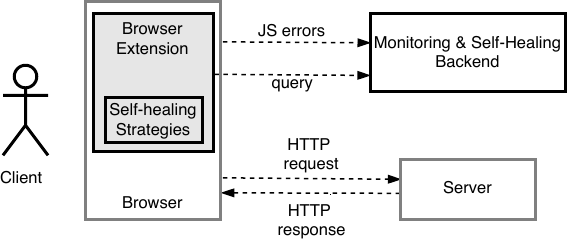}
\caption{The architecture of \extension. The key idea is that the browser extension intercepts the HTTP requests of the server and applies a self-healing strategy when an error is known in the request directly in the browser.}
\label{fig:architecture_extension}
\end{figure}

\autoref{fig:architecture_extension} presents the architecture of \extension.
It is composed of two main parts.
\vspace{-1.5em}
\begin{enumerate}
\item \extension (see \autoref{sec:extension}) is a browser extension that intercepts the HTML and Javascript requests between the browser and page rendering. 
{The extension is also responsible for applying} the self-healing strategies that handle Javascript errors. 
\item The Monitoring Backend (see \autoref{sec:backend_extension}) stores information about the known errors that have happened and the success statistics of each self-healing strategy for each error. This part can be share with the Monitoring and Self-healing Backend of \tool.
\end{enumerate}

\subsubsection{The Extension}\label{sec:extension}

The extension part of \extension contains all the logic required to apply the self-healing strategies to automatically handle Javascript errors directly inside the browser.

\autoref{algo:extension} shows the workflow of \extension.
\extension listens to all web requests for each tab in the browser (\autoref{algo:bikiniext:request}).
For each main request of a tab (the request that corresponds to the URL of the bab), \extension requests the backend to know which Javascript resource has thrown an error in the past (\autoref{algo:bikiniext:errors}) and it injects a monitoring script in the tab (\autoref{algo:bikiniext:inject}).
When an error happens on the client browser, it is sent to \tool's backend for being saved in a database.

For all the other requests, \extension checks if they triggered errors in the past (\autoref{algo:bikiniext:check}).
If it is the case, it requests the response of the request to the Web Server (\autoref{algo:bikiniext:response}).
Then, it applies the self-healing strategies on the requested content (\autoref{algo:bikiniext:rewrite}).
\extension redirects the request to data format URL (\autoref{algo:bikiniext:send}) that allows to send content to the browser in the base64 format, for example,  \texttt{data:text/javascript,<base64>}.

\begin{algorithm}[t]
  \begin{algorithmic}[1]
    \REQUIRE{T: a browser tab}
    \REQUIRE{W: the Web Server}
    \REQUIRE{R: self-healing strategies}
    \REQUIRE{D: \tool Backend}
    \WHILE{new $request$ to T}\label{algo:bikiniext:request}
        \IF{$request$ is main request}
            \STATE{$errors$ $\leftarrow$ $D.previous\_errors\_from(request_{url})$}\label{algo:bikiniext:errors}
            \STATE{$inject\_bikiniproxy\_code$ in T}\label{algo:bikiniext:inject}
        \ENDIF
        
        \IF{$request$ triggers an error from $errors$}\label{algo:bikiniext:check}
            \STATE{$response$ $\leftarrow$ W($request$)} \label{algo:bikiniext:response}
            \FOR{$r$ in $R$}
                \IF{$isToApply(r, errors, request, response)$}
                    \STATE{$response \leftarrow r.rewrite(response, request, errors)$}\label{algo:bikiniext:rewrite}
                \ENDIF
            \ENDFOR
            \STATE{$redirect(response)$}\label{algo:bikiniext:send}
        \ENDIF
    \ENDWHILE
  \end{algorithmic}
  \caption{The main \extension algorithm}
  \label{algo:extension}
\end{algorithm}

\subsubsection{Interface}\label{sec:interface_extension}

\begin{figure}[t]
    \centering
    \includegraphics[scale=0.5]{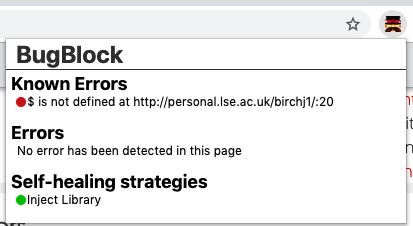}
    \caption{The user interface of \extension. It shows the known errors, the errors that are currently detected, and the self-healing strategies that are applied.}
    \label{fig:interface}
\end{figure}

{\extension is a browser extension that has a similar user interface to ad-blockers. It is directly integrated into the browser, displays its logo next to the URL bar, and it provides feedback to the users about what has been detected and what code transformations have been made and executed.}

\autoref{fig:interface} presents a screen capture of the user interface.
It provides the list of known errors for the page, the list of errors that are currently faced, and the list of the self-healing strategies that have been applied to the current page.
The interface can be extended to provide further statistics to the users, such as the most frequent errors and the websites that contain the biggest number of errors per visited page.
Those statistics can help to educate the user to know what is happening on their web session.

\subsubsection{Monitoring and Self-healing Backend}\label{sec:backend_extension}

{
The Monitoring and Self-healing Backend of \extension is the same component as the one present in \tool. 
If some users use the proxy, while others use the browser extension, having the same component allows our system to share the self-healing knowledge between the two groups of users. 
In theory, this component could also be directly integrated into the browser extension.
This would allow us to have a complete decentralized approach and to increase the privacy since no content will ever be sent to an external server.
}

\subsection{Difference between \tool and \extension}

\begin{table}[t]
    \centering
    \begin{tabularx}{\columnwidth}{l|l|X}
\textbf{Characteristics} & \textbf{\tool} & \textbf{\extension} \\\hline
Environment & HTTP proxy & Browser extension \\
Difficulty of installation & Specialist & {End-user} \\
Permission level & Full control of HTTP traffic & Limited by the browser API\\
Feedback level  & Developer feedback via a dashboard & User feedback via the extension and developer feedback via a dashboard \\
\# self-healing strategies & 5 & 5 \\
Self-healing location & Server-side & Client-side \\
Update & Handle by the server owner & Handle automatically by the extension store \\
    \end{tabularx}
    \caption{Summary of the differences between \tool and \extension.}
    \label{tab:difference}
\end{table}

\tool and \extension are two different implementations of the same approach. They aim to provide self-healing ability to web applications.
\tool is an HTTP proxy, and \extension is a browser extension.
Both tools intercept HTTP requests to inject self-healing strategies to handle known Javascript errors.

Despite those similarities, \tool and \extension are different and have different characteristics.
\autoref{tab:difference} highlights their differences.
The proxy-based approach of \tool is the most powerful approach of the two:
it allows to modify all the requests that go through the proxy freely.
However, it is complex to be set up by end-users and introduces a potential security issue since \tool behaves as a man in the middle.

On the other hand, the main strength of \extension is that it is easy to be set up by end-users.
This set up also contributes to limit the performance overhead since all the modifications are made locally.
There is no centralized server that needs to handle all the requests from all potential users of the system.
However, \extension is limited by the API of the browsers. Therefore it is not possible to modify the scripts that are present in the main HTML page of the website, and consequently limits the ability of the tool.
An additional advantage of \extension over \tool is that it provides a direct feedback to the users, i.e., to tell her whether a self-healing strategy has been applied (see \autoref{sec:interface_extension}).
{The final advantage of \extension is that the implementation is less complex compared to \tool. This difference in complexity makes \extension a more reliable approach compared to \tool.
The technical difference between the two implementations is described in \autoref{sec:implementation}.}

{In the evaluation (see \autoref{sec:evaluation}), we compare the effectiveness of \tool compared to \extension.
The goal is to identify how much is the decrease in effectiveness with the extension-based approach.
It is important to analyze this to understand the trade off between applicability and performance.}

\subsection{Implementation}\label{sec:implementation}

{
In this section, we describe the technical aspects of \tool and \extension.
The source code and the usage examples are publicly available on GitHub \cite{repo}.

\subsubsection{Implementation of \tool}

The implementation of \tool is composed of two main parts: the proxy itself and the code rewriting part.

\paragraph{Implementation of the Proxy}

As previously explained, \tool is a technique that intercepts the requests between the clients and the server and modifies them on the fly.
This is also known as an HTTP-proxy or a Man-in-the-Middle technique \cite{callegati2009man} depending on the usage.
\review{Since \tool is designed as a system or browser proxy not as a server proxy, it implies that \tool will serve content from different servers.}
This is actually close to a man-in-the-middle technique since it needs to intercepts HTTPS requests that are encrypted between the clients and the web servers.
In order to decrypt them, we need to perform a man-in-the-middle certificate spoofing. It consists of installing a root certificate in the browser. 
This root certificate is then used to decrypt all requests between the client and the webserver.
The proxy then modifies the request, re-encrypts it and finally sends it the webserver.
We base the implementation of the proxy on an existing proxy AnyProxy.\footnote{anyproxy Github repository: \url{https://github.com/Alibaba/anyproxy}} AnyProxy is a monitoring proxy designed by Alibaba to assist the debugging of their web systems.
We modify AnyProxy to be able to modify the requests, and we include the system that allows us to easily add new self-healing strategies.

For the evaluation, the proxy is combined with puppeteer\footnote{puppeteer GitHub repository: \url{https://github.com/GoogleChrome/puppeteer}} in order to automate the process.

\paragraph{Implementation of Code Rewriting}

The second component is responsible for rewriting the source code of the web pages.
This poses four main challenges:
1) identify the main request giving the HTML that defines header and body of the page,
2) regroup all the requests from the same session
3) localize the embedded Javascript scripts in HTML source code
4) being fast enough in order not to disturb the user experience.

The two first challenges are related to track the requests and to know which requests need to be rewritten.
Indeed, in order to monitor the Javascript errors, \tool needs to inject a monitoring script in all the web pages. 
The only way to inject this script is to identify the main request of the page and to inject the monitoring script inside it.
The naive approach of injecting the monitoring script in all HTML requests does not work. Indeed, some HTML requests cannot contain Javascript code, and therefore, this strategy would introduce additional bugs in the web application under consideration.
Our solution is to only inject the script in HTML requests that have a HEAD HTML tag. The drawback of this solution is an increase of required processing, because it requires to analyze all HTML responses.

The second challenge is to link all the Javascript resources to the main page, for example, the page \texttt{foo.com} loads the script \texttt{bar.com/jquery.js} and we need to create a logical link between that resource and the main page \texttt{foo.com}.
This is required in order to be able to track down which resource contains the bug, or which page contains a bug.
In order to handle this challenge, the proxy defines a unique ID on each main page. This ID is then used to \review{track} the Javascript resources that are loaded on each page.

The third challenge is to rewrite the Javascript that is embedded directly in the HTML. This is a problem since we rely on the line number defined in the Javascript error to rewrite the Javascript.  
Since the content of the HTML page can be modified dynamically in Javascript, the line number of the error can be impacted which would break the causal relationship between line numbers and errors.
We handle this problem by 1) looking at the exact location where the error has been triggered and 2) verifying that the surrounding lines match the error by looking at the variable names and function name.

The final challenge is about the performance of source code rewriting.
Parsing and iterating over the HTML/Javascript AST is CPU intensive.
Therefore, the self-healing strategies have to be optimized to reduce the number of parsed AST and the number of times the AST is traversed.
For this, we designed the plugin system for the self-healing strategies in a way that allows to only parse and print once each Javascript resource.
This drastically increases performance.

We use htmlparser2\footnote{htmlparser2 Github repository: \url{https://github.com/fb55/htmlparser2}} to parse and iterate the HTML AST and the library babel.js\footnote{babel.js GitHub repository: \url{https://github.com/babel/babel}} for analyzing and transforming the Javascript abstract syntax tree.
The implementation of \tool is composed of 4378 lines of Javascript code and 17 dependencies (504.804 lines of code).

\subsubsection{Implementation of \extension}

In this section, we present the prototype implementation of \extension. 
\review{\extension does not have challenge 1, 2 and 4 described in the implementation of \extension.}
With a browser extension, we can directly know from which web page a request comes and what the main request is. Therefore, challenge 1 and 2 do not exist. 
For challenge 4, \extension is less impacted by the performance problem because, firstly, it does not need to parse the main request to inject the monitoring script. Secondly, the AST parsing and iterating are directly executed  by the client.
Therefore a single server does not need to handle all the load because the load is distributed among all clients.

However, \extension suffers from a different challenge. Browsers do not provide a direct API to modify Javascript code, which we require to apply the self-healing strategies. 
The solution that we implemented consists of blocking  the request that loads the resource, and then
create a new HTTP request using the Ajax API that downloads the Javascript resource as a textual file.
Once the file is downloaded, it can be rewritten and injected back to the web page in order to be executed.

The implementation of \extension is composed of 1634 lines of Javascript code and four dependencies (39866 line of code).}

\section{Evaluation}\label{sec:evaluation}

In our evaluation, we answer the following research questions.

\newcommand\rqRepairability{\question{1}{[Effectiveness] How are \tool and \extension effective at automatically fixing Javascript errors in production, without any user or developer involvement?}\xspace}
\rqRepairability
The first research question studies if it is possible to handle field Javascript errors with our proxy-based approach.
We will answer this question by showing how real-world errors have been handled with one of our implemented self-healing strategies.

\newcommand\rqImpact{\question{2}{[Outcome] What is the outcome of self-healing strategies with \tool and \extension on the page beyond making the error disappear?}\xspace}
\rqImpact
In this research question, we explore what are the possible outcomes of \tool on buggy web pages.
We will answer this research question by presenting real-world case studies of different possible outcomes.

\newcommand\rqEffectiveness{\question{3}{[Comparison] Do the different self-healing strategies perform equivalently?}\xspace}
\rqEffectiveness 
We present to what extent the different self-healing strategies are used:
Which type of errors are handled by the five self-healing strategies?

\subsection{Experimentation Protocol}

We set up the following experimentation protocol to evaluate \tool and \extension.
Our idea is to compare the behavior of an erroneous web page, against the behavior of the same, but self-healed page using our tools: \tool and \extension.
In order to achieve this goal, we apply the experimentation protocol twice, one for each tool.
The comparison is made at the level of ``web trace'', a concept we introduce in this paper, defined as follows.

\begin{definition}[\trace]
A \trace is the loading sequence and rendering result of a web page.
A \trace contains 
1) the URL of the page
2) all the resources (URL, content,  see Definition \ref{def:resource})
3) all the Javascript errors that are triggered when executing the Javascript resources
and 
4) a screenshot of the page at the end of loading.
\end{definition}

Given a benchmark of web pages with Javascript errors, the following steps are made.
The first step is to collect the \trace of each erroneous web page.
The second step is to collect the new \trace of each erroneous web page with one of our implementation of proxy-based failure-oblivious approaches (Recall that all resources are rewritten by our five self-healing strategies).
In addition to the \trace, we also collect data about the self-healing process:
the strategies that have been activated, defined by the tuple (initial error, strategy type).
The third step is to compare for each web page the original \trace against the self-healed \trace.
The goal of the comparison is to identify whether our approach is able to heal the Javascript errors. For instance, the comparison may yield that all errors have disappeared, that is a full self-healing.

We apply this protocol twice, once for \tool and once for \extension. 
At the end of the experimentation, we have for each tool the \trace and the self-healing strategies that have been applied for each bug of the benchmark.

\subsection{Construction of a Benchmark of Javascript Field Errors}\label{sec:benchmark}

To evaluate our approaches, we need real-world Javascript that are reproducible.
For each reproducible errors, we want to compare the behavior of the web page with and without the self-healing approaches.
To our knowledge, there is no publicly available benchmark of reproducible Javascript errors.
We create a new benchmark. We call it the \bench benchmark.
The creation of our benchmark is composed of the following steps:
\begin{enumerate}
\item Randomly browses the web to discover web pages on Internet that have errors (see \autoref{sec:url_finder}).
\item Collect the errors and their execution traces (see \autoref{sec:state_collector}).
\item Ensure that one is able to reproduce the errors in a closed environment(see \autoref{sec:reproduction_proxy}).
\end{enumerate}

\subsubsection{Web Page Finder}\label{sec:url_finder}

The first step of the creation of \bench is finding web pages that contain errors.
In order to have a representative picture of errors on the Internet, we use a random approach.
Our methodology is to take randomly two words from the English dictionary and to combine those two words in a Google search request.
A fake crawler then opens the first link that Google provides. If an error is detected on this page, the page URL is kept as tentative for the next step. 
The pros and cons of this methodology are discussed in \autoref{sec:threats_validity}

\begin{table}[t]
\centering
\caption{Descriptive Statistics of \bench}
\label{tab:main_stat}
\begin{tabularx}{0.9\columnwidth}{X|r}
\hline
Crawling stats & Value \\ \hline
\# Visited Pages & \visitedPage \\
\# Pages with Error & \nbBug (4.5\%)\\
\hline
Benchmarks stats & Value \\ \hline
\# Pages with Reproduced Errors & \nbReproducedBug \\
\# Domains & 466 \\

\# Average \# resources per page  & 102.55 \\
\# Average scripts per page & 35.51 \\
\# Min errors per page & 1 \\
\# Average errors per page & 1.49 \\
\# Max errors per page & 10 \\
\# Average pages size  & 1.98mb \\
\end{tabularx}

\end{table}

\begin{figure}[t]
\centering
\includegraphics[width=\columnwidth]{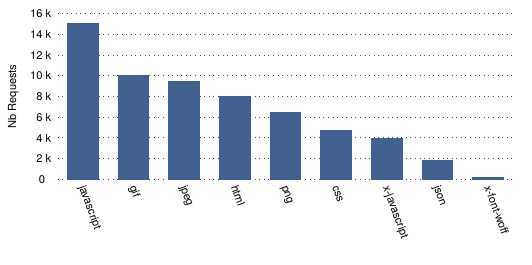}
\caption{Bar plot of the number of requests by content-type.}
\label{fig:contenttype}
\end{figure}

\subsubsection{Web Trace Collector} \label{sec:state_collector}

The Javascript environment is highly dynamic and asynchronous.
It means that many errors are transient and as such are not reproducible in the future, even in a very short period of time after their observation.

For identifying \review{reproducible} errors, our idea is to collect the \trace of the erroneous page and to try to reproduce the exact same \trace in a controlled environment, see \autoref{sec:reproduction_proxy}

We implement the trace collection using the library puppeteer from Google\footnote{puppeteer repository \url{https://github.com/google/puppeteer}}, which provides an API to control Chrome in a programmatic manner. The big advantage of this library is that it uses the same browser engine as Chrome end-users, meaning that, by construction, \bench is only composed of errors that really happen on user browsers. 

Since Javascript is mostly asynchronous, the Web Trace Collector waits for the end of loading where loading is defined as follows:
1) it opens the URL,
2) it waits for seven seconds, in order to load and execute all resources, in particular, Javascript files. 
3) it scrolls the page to the bottom, in order to trigger additional initialization and Javascript execution.
4) it waits again for one second.

During this process, the Web Trace Collector logs 1) all errors that occur in the browser console and 
2) all the requests (including the HTTP headers and the body) made from the browser.
When the page is completely loaded, a screenshot of the page is taken, it provides a visual representation of the page.
At the end of this process, for each page, the collected data is stored on disk if at least one error has been logged during the page browsing.

\subsubsection{Web Page Reproduction}\label{sec:reproduction_proxy}

The last step of the benchmark creation consists of verifying that the collected errors can be reproduced.
We consider that we succeed to reproduce the behavior of the web page when the observed errors during reproduction are identical to the ones in the originally collected \trace.

The reproduction of the error is done by browsing the erroneous page again, but instead of using the resources from the Internet, the Web Page Reproduction is cut from the Internet and only serves the resources stored on disk.
In addition, it denies all the requests that have not been observed during the initial collection of the page.

\begin{table}[t]
\centering
\caption{The Top 10 Error Types in \bench (left-hand side).}
\label{tab:errors_handled}
\begin{tabularx}{\linewidth}{r|X|r|r|r}
\# & Error messages
& \# Web Pages & \# Domains
& {\tabincell{c}{\# Initial Errors}}\\ \hline
1 & XXX is not defined & 200 & 166 & 307 \\
2 & Cannot read property XXX of null & 156 & 126 & 176 \\
3 & XXX is not a function & 92 & 86 & 111 \\
4 & Unexpected token X & 54 & 51 & 61 \\
5 & Cannot set property XXX of null & 21 & 17 & 24 \\
6 & Invalid or unexpected token & 18 & 12 & 21 \\
7 & Unexpected identifier & 13 & 11 & 15 \\
8 & Script error for:  XXX & 8 & 3 & 10 \\
9 & The manifest specifies content that cannot be displayed on this browser / platform. & 5 & 5 & 7 \\
10 & adsbygoogle.push() error: No slot  & 4 & 4 & 7 \\\hline
  & 53 different errors & 555 & 466 & 826  \\
\end{tabularx}
\end{table}

\begin{table}[t]
\centering
\caption{The effectiveness of \tool and \extension (right-hand side).}
\label{tab:results}
\begin{tabularx}{\linewidth}{r|X|r|r|r|r}
  &  &  \multicolumn{2}{c|}{\tool} &    \multicolumn{2}{c}{\extension} \\ \cline{1-6}
\multirow{-2}{*}{\#} & \multirow{-2}{*}{Error messages} & \#Healed Errors & Improvement & \#Healed Errors & Improvement \\ \hline
1 & XXX is not defined &  184 & 59.93\% & 36 & 11.72\% \\
2 & Cannot read property XXX of null &  42 & 23.86\% & 10 & 5.74\% \\
3 & XXX is not a function &  11 & 9.9\% & 20 & 18.01\% \\
4 & Unexpected token X & 2 & 3.27\%  & 8 & 13.11\% \\
5 & Cannot set property XXX of null & 11 & 45.83\% & 0 & 0\% \\
6 & Invalid or unexpected token &  0 & 0\% & 0 & 0\% \\
7 & Unexpected identifier & 0 & 0\% & 0 & 0\% \\
8 & Script error for:  XXX & 2 & 20\% & 0 & 0\% \\
9 & The manifest specifies content that cannot be displayed on this browser / platform. & 0 & 0\% & 0 & 0\% \\
10 & adsbygoogle.push() error: No slot  & 0 & 0\% & 0 & 0\% \\\hline
  & 53 different errors & 248/826 & 30.02\% & 88/826 & 10.67\% \\
\end{tabularx}
\end{table}

\subsection{Description of \bench}

\autoref{tab:main_stat} gives the main statistics of \bench.
The Web Page Finder visited a total of \visitedPage pages, and \nbBug of the pages contains at least one error (4.5\%), out of which \nbReproducedBug errors have been successfully reproduced.
The final dataset contains errors from 466 different URL domains representing a large diversity of websites.
There is, on average, 1.49 error per page, and each page has between one and ten errors.

\autoref{tab:errors_handled} presents the top 10 of the errors present in \bench.
In total \bench contains 53 different error types for a total of 826 collected errors.
69\% of the Javascript errors are the first three error types: \mycode{XXX is not defined}, \mycode{Cannot read property XXX of null} and \mycode{XXX is not a function}.
\autoref{fig:contenttype} presents the number of requests for the top 9 resource types.
In our benchmark, the most common external resources are Javascript files.
The rest of the distribution illustrates how complex modern web pages are.
For sake open of open-science, \bench and its mining framework are available on Github \cite{repo}. 

\subsection{RQ1: Effectiveness of Self-healing Web Applications}

We now present the results of the first research question.
\autoref{tab:results} shows the top 10 types of errors in the considered benchmark and how they are handled by \tool and \extension.

The first column contains the rank of the error type.
The second column contains the error type, represented by the message of the error.
The third column contains the number of healed errors with \tool.
The fourth column contains the percentage of errors fixed with \tool.
The fifth column contains the number of healed errors with \extension.
The sixth column contains the percentage of errors fixed with \extension.

The first major result lies in the first row.
It presents the error ``XXX is not defined'', which is the most common on the web according to our sampling.
This error is present in 200 web pages across 166 different domains (see \autoref{tab:errors_handled}).
It is thrown 307 times, meaning that some web pages throw it several times.
With \tool, this error is healed 184/307 times, which represents a major improvement of 59.93\%.
With \extension, this error is healed 36/307 times, which represents a major improvement of 11.72\%.

A second major result is that \tool is able to handle at least one error for the five most frequent Javascript errors.
It succeeds to heal between 3.27\% and 59.93\% of the five most frequent Javascript errors in our benchmark.
Overall, \tool handles 248 errors, and \extension handles 88 errors.
It means that \tool reduces by 30.02\% the number of errors in the benchmark and \extension 10.67\%.
{Those results also indicate that the drawback of using a browser extension is almost 20\% fewer handled error.}

Now we discuss the categories of healed errors.
We identify whether:
\begin{enumerate}
\item \emph{All errors disappeared:} no error happens anymore in the page loaded with our tool, meaning that one or a combination of rewriting strategies have removed the errors.
\item \emph{Some errors disappear:} there are fewer errors than in the original \trace.
\item \emph{Different errors appear:} at least one error still, and it is a new error (new error type or new error location) that has never been seen before.
\item \emph{No strategy applied:} the error type is not handled by any of the strategies, and thus there are the same errors than in the original \trace.
\end{enumerate}

\begin{table}[t]
\centering
\caption{Analysis of the healing effectiveness per page.}
\label{tab:repairability}
\begin{tabularx}{0.97\columnwidth}{X|r|r|r|r}
                             &    \multicolumn{2}{c|}{\tool} &    \multicolumn{2}{c}{\extension} \\ \cline{1-5}
\multirow{-2}{*}{Metric Name}& \# Pages & Percent & \# Pages & Percent\\ \hline
All Errors Disappeared       & 176/\nbReproducedBug & 31.76\% & 87/\nbReproducedBug & 15.67\% \\
Some Errors Disappeared      & 42/\nbReproducedBug & 7.58\%   & 9/\nbReproducedBug  & 1.62\% \\
Different/Additional Errors  & 140/\nbReproducedBug & 25.27\% & 52/\nbReproducedBug & 9.37\% \\
\hline
No Strategy Applied          & 196/\nbReproducedBug & 35.31\% & 407/\nbReproducedBug & 73.15\% \\
\end{tabularx}
\end{table}

\autoref{tab:repairability} presents the number of web pages per category.
The first line of \autoref{tab:repairability} shows that the number of web pages that have all the Javascript errors healed by \tool and \extension.
\tool is able to handle all errors for 176/\nbReproducedBug (31.76\%) web pages of the \bench benchmark.
\extension is able to handle all errors for 87/\nbReproducedBug (15.67\%) of the \bench benchmark.
The second line shows the number of web pages that have been partially self-healed, by partially, we mean that the number of Javascript errors decrease but are still not zero.
With \tool, 42 web pages contain fewer errors than before, and with \extension, nine web pages are in this case.
The third line shows the number of web pages that have new errors than before: 
with \tool, this case is detected for 140/\nbReproducedBug (25.27\%) web pages, and 52/\nbReproducedBug (9.42\%) with \extension.
The last line shows the number of web pages where none of the strategies has been applied:
in 196/\nbReproducedBug (35.13\%) the errors are of a type that is not considered by \tool.
In the case of \extension, 406/\nbReproducedBug (73.28\%) of the buggy pages have not been handled by \extension.
To better understand the case where no strategy can be applied, we perform a manual qualitative analysis.

\begin{figure*}[t]
\centering
\begin{subfigure}{.5\textwidth}
  \centering
  \includegraphics[trim={0 50cm 0 5cm},clip,width=0.97\linewidth]{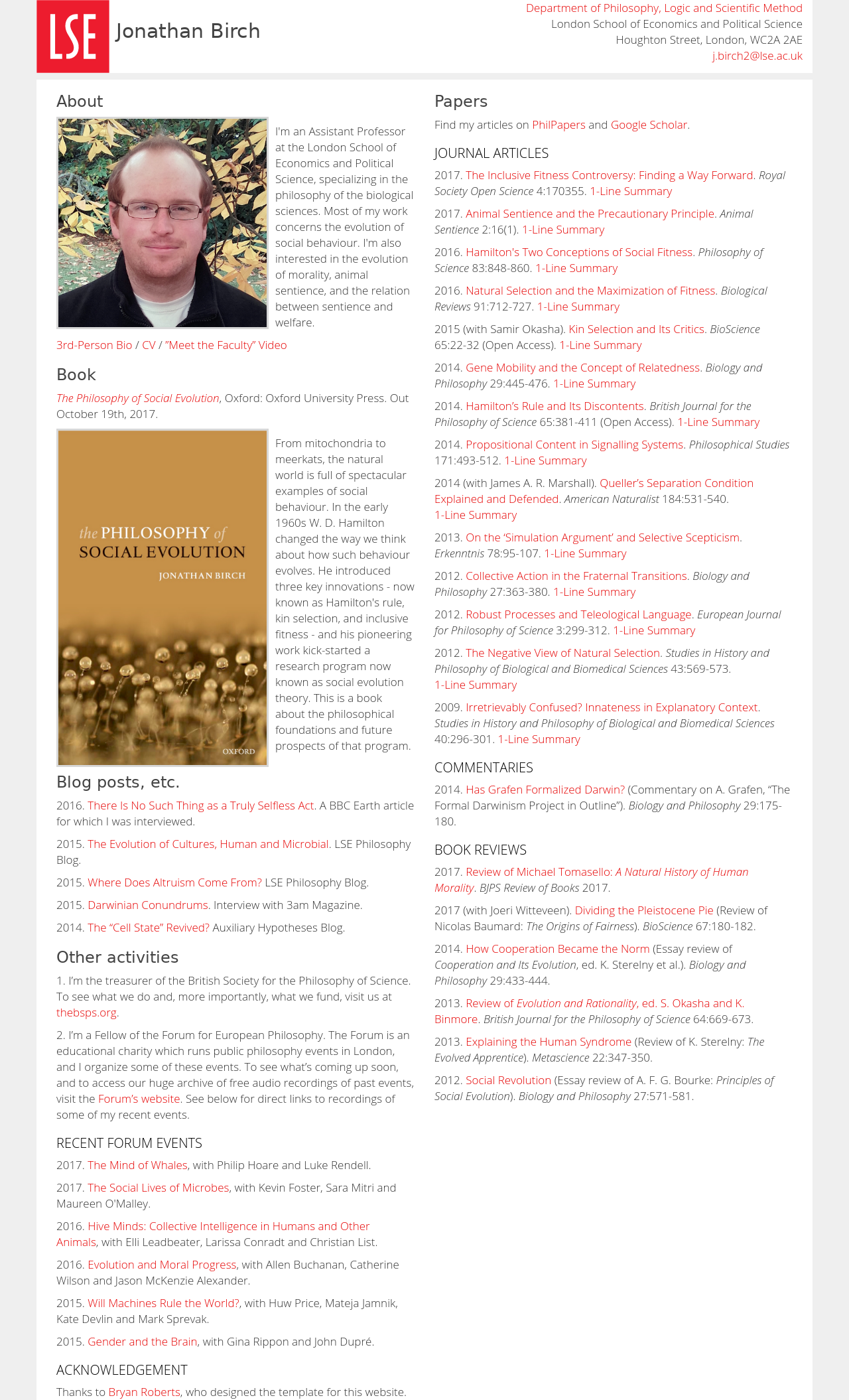}
  \caption{Without \tool, some content is missing}
  \label{fig:sub1}
\end{subfigure}\begin{subfigure}{.5\textwidth}
  \centering
  \includegraphics[trim={0 69.5cm 0 5cm},clip,width=0.97\linewidth]{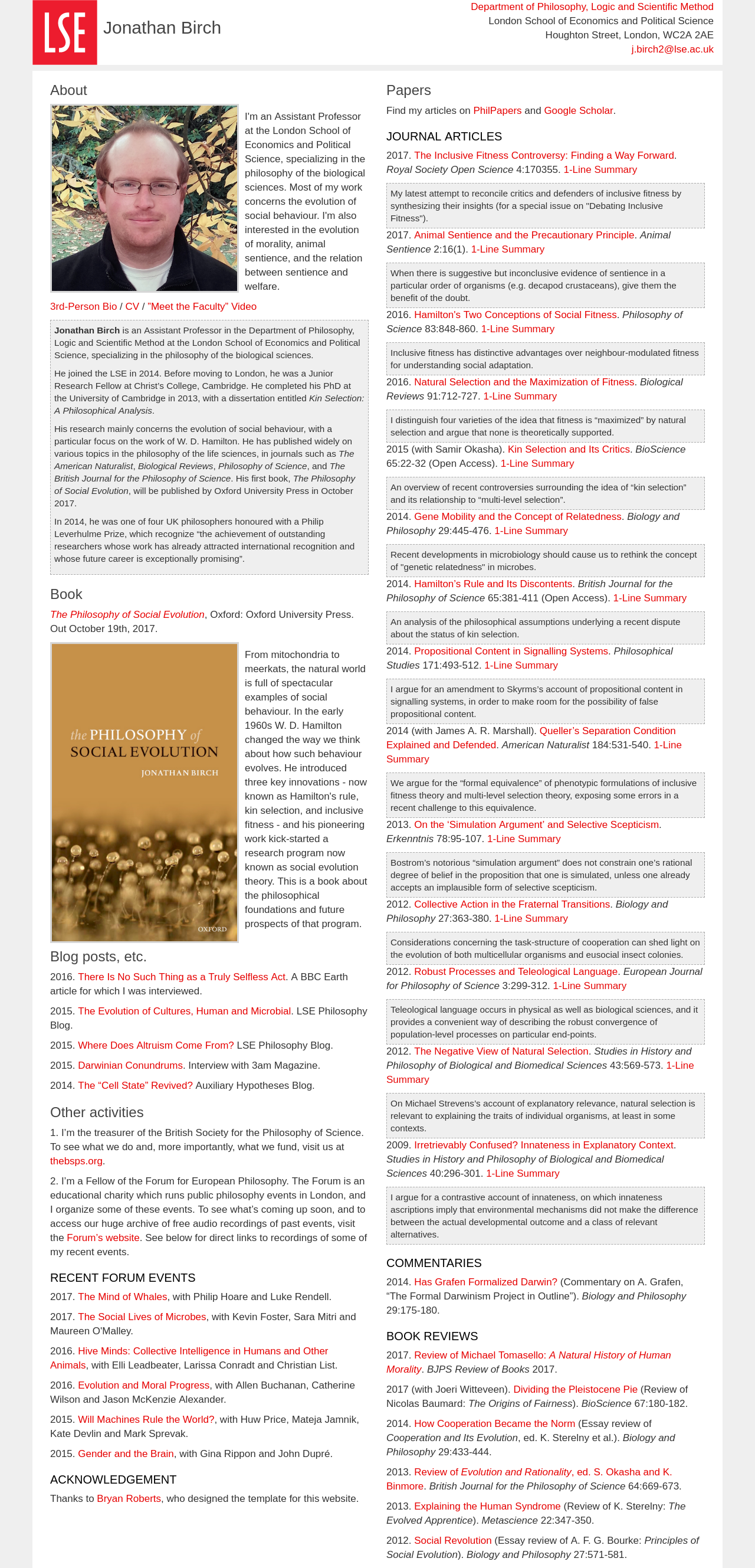}
  \caption{Using \tool, the page loading is self-healed.}
  \label{fig:sub2}
\end{subfigure}
\caption{A real web page suffering from a Javascript bug. With \tool, the bug is automatically healed, resulting in additional information provided to the web page visitor.}
\label{fig:diff}
\end{figure*}

\begin{figure}[t]
\centering
\includegraphics[width=\columnwidth]{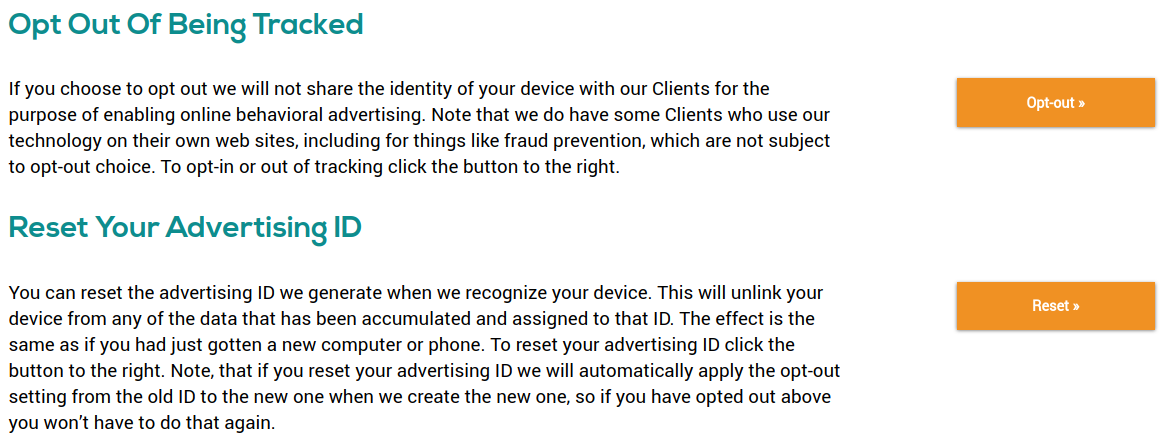}
\caption{The two buttons in orange are missing in the original buggy page. When \tool is enabled, the two orange buttons provide the user with new user-interface features.}
\label{fig:new_feature}
\end{figure}

\paragraph{Unhandled Errors}\label{sec:unhandled_errors}
The errors are unhandled when none of the five rewriting strategies succeed to heal the errors.
In our experiment, this type of scenario is frequent for \extension with 73.28\% of the web pages not being handled, and in a lower proportion for \tool with 196/\nbReproducedBug of web pages have errors not healed, which represents 35.13\% of the erroneous web pages of \bench.

The difference between the number of unhandled errors between \tool and \extension is related to a technical limitation of \extension.
\extension is not able to rewrite the HTML content of the page before they are executed.
It means that in all the cases where the error is triggered inside an HTML page, for example, in a \textless SCRIPT\textgreater tag, \extension is not able to handle the error.
This is due to a technical limitation of browser extensions.

For the other \review{cases}, we identify two main root causes.
The first cause of non-healed errors is that the error type is not supported.
For example, the web page \url{http://dnd.wizards.com/articles/unearthed-arcana/artificer} is loading a JSON file.
However, the JSON file is invalid, and the browser does not succeed to parse it which produces an \mycode{Unexpected token <} error.
None of the five strategies is able to handle malformed JSON errors.
The second cause of non-healed errors is that the self-healing strategies have not enough information to rewrite the resource.
For example, the web page \url{http://moreas.blog.lemonde.fr/2007/02/28/le-pistolet-sig-sauer-est-il-adapte-a-la-police/} contains the error \mycode{Cannot read property 'parents' of undefined}, this error should be healed with ``Object Creator'' rewriting.
However, the trace of the error does not contain the URL of the resource that triggers this error because the Javascript code has been unloaded.
Consequently, ``Object Creator'' is not able to know which resource has to be rewritten to handle the error.

In summary, \autoref{tab:repairability} shows that \tool is almost able to heal all the errors from a third of \bench.
The second third of the benchmark is pages that cannot be healed with \tool.
The last third contains web pages that are partially headed or that the self-healing strategies produce new errors.
In the case of \extension,  \autoref{tab:repairability} shows that \extension is not able to handle the errors in the majority of the cases (73\%) and is able to handle the errors of 15.67\% of the pages completely.
It shows that the technical limitations of browser extension have an important impact on the healing effectiveness of \extension.

\answer{1}{
\textbf{How are \tool and \extension effective at automatically fixing Javascript errors in production, without any user or developer involvement?}
\tool is effective the handle the five most frequent Javascript errors present in our benchmark. With the currently implemented self-healing rewriting strategies, \tool is able to fully heal 248/826 (30.02\%) of all errors, representing 196/555 (31.76\%) of all buggy web pages of our benchmark.
The healing effectiveness of \extension is reduced by the technical limitation of browser extensions.
Our experiment shows that 10.67\% of the errors have been fully healed by \extension.
{This shows that despite being more practical, a browser extension has a lower performance.}}

\subsection{RQ2: Outcome}

In this second research question, we focus on category ``All Errors Disappeared'', and further refines the classification as follows:
\begin{enumerate}
\item The errors have disappeared, but the end-user can see no behavioral change.
\item The errors have disappeared, and new UI features (e.g., new buttons) are available to the end-user.
\item The errors have disappeared, and new content is available for the end-user.
\end{enumerate}
Contrary to RQ1, it is not possible to automatically classify all pages with this refined category, because it requires a human-based assessment of what is new content or new features. For this reason, we answer this RQ with a qualitative case study analysis, and we do not consider \tool and \extension separately since the case studies are valid for both tools.

\subsubsection{Error Handled but No Behavior Change}

A healed error does not automatically result in a behavior change in the application.
For example, this is the case for the website \url{https://cheapbotsdonequick.com/source/bethebot}, which triggers the error \mycode{"module" is not defined}.
This error is triggered by line \mycode{module.exports = tracery;}.
This type of line is used to make a library usable by another file in a Node.js environment.
However, Node.js has a different runtime from a browser, and the module object is not present, resulting in the error.
With \tool, the self-healing strategy ``Object Creator'' automatically initializes the variable \mycode{module}, however, since this line is the last line of the executed Javascript file, this has absolutely no further consequence on the execution or the page rendering.
This means that the error was irrelevant.
However, from a self-healing perspective, this cannot be known in advance.
From a self-healing engineering perspective, the takeaway is that it is more straightforward to heal irrelevant errors than to try to predict their severity in advance.

\subsubsection{New Feature Available}

One possible outcome of our approach is that the self-healing strategy unlocks new features.
For example, this is the case of \url{https://bluecava.com/}.
This page has an error, shown in \autoref{lst:error_feature}, which is triggered because the developer directly accesses the content of Ajax requests without checking the status of the request.
However, there are requests that are denied due to cross-domain access restrictions implemented in all browsers.
Since the developer did not verify if there is an error before accessing the property 'id' on a null variable, the Javascript event loop crashes.

With our approach, the self-healing strategy ``Object Creator'' ensures the initialization of the variable if it is null.
This execution modification allows the execution to continue and to finally enter into an error handling block written by the developer, meaning that the event loop does not crash anymore.
The execution of the page continues and results in two buttons being displayed and enabled for the end-user.
\autoref{fig:new_feature} presents the two buttons that are now available for the user.

\begin{lstlisting}[caption={Error on the web page \url{https://bluecava.com/}}, label=lst:error_feature, float]
Uncaught TypeError: Cannot read property 'id' of null
    at bluecava.js?v=1.6:284 ...
    at post (bluecava.js?v=1.6:40)
    at identify (bluecava.js?v=1.6:156) ...
\end{lstlisting}

\subsubsection{New Content Available}

One other outcome of our approach is that additional content is displayed to the end-user.

Let us consider the web page \texttt{http://personal.lse.ac.uk/birchj1/} that is the personal page of a researcher. This page triggers the following error: \mycode{\$ is not defined at (index):20}

This error is thrown because a script in the HTML page calls the jQuery library before the library is loaded.
The script that throws the error is responsible for changing the visibility of some content on the page. Consequently, because of the error, this content stays hidden for all visitors of the page.

Using our tool, the error is detected as being caused by a missing jQuery library.
This error is healed by rewriting strategy ``Library Injector''
Consequently, the missing the jQuery library is injected in the buggy page.
When the rewritten web page is executed, jQuery is available, and consequently, the script is able to change the visibility of hidden HTML elements, resulting in newly visible content.

\autoref{fig:diff} presents the visual difference between the original page (left side), and loaded with \tool (right side).
All the elements in a grey box on the right-hand side are missing on the left image. They have appeared thanks to self-healing.

Finally, we have manually checked the presence of potentially harmful effects. By manually analyzing
a random sample of 25 self-healed subjects, we did not find a single harmful effect.

\answer{2}{
\textbf{What is the outcome of self-healing strategies with \tool and \extension on the page beyond making the error disappear?}
We observe three outcomes in our benchmark: (1) no visible change; (2) new features; and (3) new content. 
\tool and \extension are able to restore broken features or broken content automatically. We have not observed any harmful effect of speculative execution.}

\subsection{RQ3: Strategies}

In this research question, we compare the five different self-healing strategies.
For each strategy, \autoref{tab:strategies_comparison} shows the number of times it has been activated to heal errors of \bench, with our two tools: \tool and \extension.
The last column presents the number of different error types for which the strategy has been selected.
For example, the first row of \autoref{tab:strategies_comparison} shows that ``Line Skipper'' has been selected to handle 233 errors with \tool and 89 with \extension, and it has healed four different error types.

\begin{table}[t]
\centering
\caption{The number of activations of each self-healing strategy and the number of error types that the strategy can handle.
}
\label{tab:strategies_comparison}
\begin{tabularx}{0.97\columnwidth}{X|r|r|r}
 & \multicolumn{2}{c|}{\# Activations} & \\\cline{1-3}
\multirow{-2}{*}{Self-healing Strategies} & \tool & \extension &
\multirow{-2}{*}{\tabincell{c}{\# Supported \\ Error Types}}\\\hline
Line Skipper          & 233 &89 & 4 \\
Object Creator        & 109 &17 & 2 \\
Library Injector      & 75  &55 & 3 \\
HTTP/HTTPS Redirector & 18  &18 & NA\\
HTML Element Creator  & 14  &11 & 2 \\
\end{tabularx}
\end{table}

In the case of \tool, the most used strategy is ``Line Skipper'' with 233 activations.
It is also the strategy that supports the highest number of different error type: 
1) ``XXX is not defined'', 2) ``XXX is not a function'', 3) ``Cannot read property XXX of null'', 4) ``Cannot set property XXX of null''.
On the other hand, \extension uses the most the strategy ``Line Skipper'' strategy with 89 activations, followed by ``Library Injector'' with 55 activations.
The second most used strategy for \tool is ``Object Creator'' with 109 errors for which it has initialized a null variable.
This strategy handles two different error types: ``Cannot set property XXX of null'' and ``Cannot read property XXX of null''.
These two strategies have something in common, they target the failure point, the symptom, and not the root cause (the root cause is actually unknown).
For example, the error \mycode{CitedRefBlocks is not defined} is triggered in the web page \url{https://www.ncbi.nlm.nih.gov/pmc/articles/PMC504719/}, because the function \mycode{CitedRefBlocks} is not defined.
Line Skipper strategy avoids the error by skipping the method call, it is a typical example of a fix at the failure point and not at the root cause of the absence of \mycode{CitedRefBlocks}.

On the contrary, ``Library Injector'' addresses the root cause of the problem: the missing library is extracted from the error message, and it is used to rewrite the content of the request. In this case, the self-healing tools exploit the fact that we have a direct relation between root cause (no included library) and symptom (unknown used library name) for this error type.

The case of ``HTTP/HTTPS Redirector'' is the opposite.
Recall that ``HTTP/HTTPS Redirector'' directly looks in the HTML body of the resource if there are scripts that will be blocked. This means that the rewriting addresses the root cause of potential future problems.
For example, the page \url{https://corporate.parrot.com/en/documents} tries to load the resource \url{http://www.google-analytics.com/urchin.js}, but the request is blocked by the browser (HTTP request in an HTTPS page).
Consequently, the Google tracking library is not loaded and function \mycode{urchinTracker} is not defined, resulting in the error \mycode{urchinTracker is not defined}. 
``HTTP/HTTPS Redirector'' strategy rewrites the URL of the resource in the \textless SCRIPT\textgreater tag to \url{https://www.google-analytics.com/urchin.js}, and this fixes the error of the page.
This strategy can potentially fix error types that we cannot envision. Hence, we do not know the exact number of handled error types, so we put ``NA'' in \autoref{tab:strategies_comparison}. 
Finally, strategy ``HTML Element Creator'' is applied to more rare errors happening only 14 times in our benchmark with \tool and 11 times with \extension. 

In this research question, we also observe that the number of activations is lower for \extension.
This observation is directly related to our observation of unhandled errors in the research question one (see \autoref{sec:unhandled_errors}).

\answer{3}{
\textbf{Do the different self-healing strategies perform equivalently?}
In our experiment, the most used strategy is ``Line Skipper'' for \tool because it is able to heal from four common error types with the same strategy.
This strategy is not the most frequent in the case of \extension because it is not able to modify the HTML page of an application.
Other self-healing strategies can be designed and added to \tool and \extension in order to address the rare error types in the long tail of field errors.}

\section{Discussion} \label{sec:discussion}

\subsection{Security Analysis}
\label{sec:security-analysis}
\tool and \extension are founded on the core failure-oblivious computing principle \cite{rinard2004enhancing}: any execution happening after the avoided failure is, in essence, speculative. This speculative execution must be sandboxed. 

The security guarantees of \tool and \extension are provided by the sandboxing in the browser and on the server-side.
First, all browsers contain very carefully engineered code to sandbox the execution of Javascript code.
This sandboxing means that 1) the Javascript code cannot access or transfer data to other tabs and windows (aka tab sandboxing)  2) the Javascript code cannot access or transfer data to other websites (cross-domain restrictions) 3) the Javascript code cannot access to the file-system.

Second, in distributed Internet applications with code running on the server-side and on the client-side, it is known that one cannot trust the execution of the client code. 
Consequently, the best practice is to protect the server-side state with appropriate checks in the REST API accessed by client-side Javascript. 
Those checks form the second sandboxing of speculative execution of this approach: the unwanted side-effects are confined to the current browser window.

{
In term of privacy, the only information that is shared between \tool, \extension, and the backend is the stacktraces of errors.
The stacktraces do not contain personal information or information that can lead to identifying a specific user or its browsing habit.
With \extension, this can even be addressed: \extension could be extended by including the backend service directly inside the browser extension. 
Using this approach, no information ever is sent by \extension to a third-party server.}

\subsection{Applicability Analysis}

The usage of \tool and \extension is practically zero cost, and as such, it is widely applicable.
First, it requires no change to the original web pages or applications.
Second, for the \tool case, the usage of an HTTP proxy in web applications is very common.
A \tool self-healing proxies can be set up by:
1) a company in front of their web content;
2) a SaaS-based provider
3) a hosting service.
\tool targets the professional sector, while \extension targets the end-user that only needs to click on one button to install the extension in its browser.
Thus, \tool and \extension target two different public.

\section{Threats to validity}
\label{sec:threats_validity}

We now discuss the threats to the validity of our experiment.
First, let us discuss internal validity. 
Our experiment is relying on the implementation of our prototype, consequently, a bug in our code may threaten the validity of our results. 
However, since the source code of the approach and of the benchmark is publicly available \cite{repo}, future researchers will be able to identify these potential bugs.
It is unknown whether the errors of our benchmark are representative of all errors in the web, and whether 96174 visited pages is enough compared to the trillions of pages of the Internet.
To our knowledge, there is no work on the \review{representativeness} of Javascript bugs.

Our approach has been carefully designed to maximize \review{representativeness}: 
1) the randomness of keyword choice allows us to discover websites about many different topics, done by a variety of persons, with different backgrounds (a website on CSS done by a web developer is likely to have fewer errors than a website on banana culture done by a hobbyist).
2) the ranking of Google for a specific query provides us with a filter which favors popular websites. If errors are detected on those websites, they likely affect many users.

\section{Related Work}\label{sec:rw}

\subsection{Javascript Error \& Repair}
Several studies on client-side Javascript been have been made by Ocariza et al \cite{ocariza2011Javascript,ocariza2013empirical,ocariza2017study}.
In 2011, they showed that most websites contain Javascript errors even in the top 100 of Alexa \cite{ocariza2011Javascript}.
They have also investigated \cite{ocariza2013empirical,ocariza2017study} the nature of the Javascript errors as follows.
They manually analyze Javascript bug reports from various web applications and Javascript libraries. 
They find that the majority of reported Javascript bugs are related to the Document Object Model (DOM).
None of this work has explored self-healing strategies for the web.

Now, we discuss the works on reproducing Javascript errors or to extract regression tests.
Wang et al. \cite{wang2017jstrace} present a technique to reproduce sequences of events that lead to a Javascript error.
Schur et al. \cite{schur2014procrawl} present a fully automatic tool to generate test scripts based on the behavior of multi-user web applications.
The goal of those works are different ours: the focuses on creating tests while we focus on healing the error on the fly, in production.

Hanam et al. \cite{hanam2016discovering} present BugAID a data mining technique for discovering common unknown bug patterns in server-side Javascript.
Hanam et al. focus on server-side bugs, on the contrary, \tool targets client-side Javascript code, and it heals the errors on the fly.

There have been several repair tools targeting Javascript front-end code.
Ocariza et al. \cite{ocariza2014vejovis} present Vejovis, a technique that suggests Javascript code modifications to handle DOM-related errors. 
Pradel et al. \cite{pradel2015typedevil} and Bae et al. \cite{bae2014safewapi} proposed tools for detecting type inconsistencies and web API misuses in Javascript, respectively.
They also present common fault types and common web API misuse patterns.
Roy et al. \cite{roy2013x} present X-PERT, an automatic technique to detect cross-browser issues in web applications.
Those works are offline program repair requiring test cases \cite{KongZWL15}, while \tool is online self-healing, in production, without the developer in the loop.

\subsection{Self-healing in Production}

There are many kinds of self-healing approaches, we refer the reader to recent surveys on this topic \cite{Monperrus2015,MarianiSurveyRepair}. 
We now concentrate on self-healing for the web.

Carzaniga et al. \cite{carzaniga2010automatic} propose a technique that automatically applies workarounds to handle API issues. The workarounds are based on a set of manually written API-specific alternative rule.
{The difference with our work is that we defined five self-healing strategies that are generic, i.e., which are not application specific. On the contrary, Carzaniga et al.'s work relies on specific templates for specific APIs and new templates have to be manually created to support new APIs.}
In subsequent work, the same group has proposed a way to automatically mine those workarounds \cite{carzaniga2015automatic}.
However, those workarounds do not consider generic Javascript errors as we do, they target API specific errors for which workarounds have been identified or mined.

Several approaches also use a proxy-based architecture in their contribution.
Kiciman et al. \cite{kiciman2007ajaxscope} present a web proxy named AjaxScope, that instruments the Javascript code to monitor the performance of web applications. It does monitoring and not self-healing.
Zhang et al. \cite{Zhang2017InteractionPF} present a technique to change the user interface of mobile applications on the fly.
In particular, they aim at providing accessible UIs to blind users.
Appelt et al. \cite{AppeltPB17} do automatic repairs of firewall rules to improve the security of web application. While a firewall and a proxy are similar, the goal and the means are different: they focus on security while we focus on availability, they change firewall rules while \tool rewrites HTML and Javascript code.

\section{Conclusion}\label{sec:conclusion}

In this paper, we have presented a novel approach to provide self-heal capabilities for the web, focusing on client-side Javascript errors and two different implementations of this approach: an HTTP proxy called \tool and a browser extension called \extension.
We have evaluated our technique on \nbReproducedBug web pages with Javascript errors, randomly collected on the Web. 
Our qualitative and quantitative evaluation has shown that \tool is effective and self-healing results in providing the web user with new features and content.
It also shown that \extension is able to heal buggy web pages but with a smaller proportion due to technical limitations of browser extensions.
Future work is required to devise new self-healing rewriting strategies for solving the maximum number of Javascript runtime errors.

\balance
\bibliography{references}

\end{document}